\documentstyle[12pt]{article}
\titlepage

\title{String Origins}
\author{Richard Atkins} 
\date{}

\begin{document}
\maketitle 
Department of Mathematics, Trinity Western University, Glover Road, Langley, 
BC V2Y 1Y1, Canada \\
\begin{abstract}
It is shown that the string concept results naturally from considerations of gravitation.
This paper describes a derivation of linearized general relativity based upon the hypotheses
of special covariance and the existence of a gravitational potential. The gravitational 
field possesses gauge invariance given by a second-order covariant derivative defining 
an associated differential geometry. The concepts of parallelism and parallel transport 
lead to string-like constructions. 
\end{abstract} 

\newpage

\section{Introduction}

Strings were originally introduced as a means of describing hadronic interactions 
\cite{gg}, \cite{hh}, \cite{kk}. After the quantum theory was found to contain the graviton 
it became a candidate for unification of the known forces  \cite{ii}, \cite{jj}. The discovery
of the anomally cancellation \cite{cc} gave further impetus to the unification programme
leading to the search for realistic models \cite{ee}, \cite{ff}.  
The character of this paradigm shift from the zero-dimensional particle, 
which had been the hypothesized elementary constituent of matter from
the time of the early Greeks until the latter part of the twentieth century, to the 
one-dimensional string has appeared ad hoc, lacking theoretical continuation from prior held 
knowledge of relativity and quantum fields. It is the purpose of this paper to
show that the string concept emanates naturally from considerations of gravitation.   

This paper presents a derivation of linearized general relativity based upon two hypotheses:  
special covariance and the existence of a gravitational potential. The approach is in close 
analogy with electrodynamics, which provides the motivation behind most of the concepts to be
introduced. The gravitational potential defines a gravitational field tensor which possesses 
gauge invariance given by second-order covariant differentiation, in contradistinction to 
first-order Yang-Mills theories. The differential geometry associated to the 
gauge transformations yields the graviton, a perturbation of the metric from a flat 
spacetime background. Moreover, the arbitrariness of the geometric structure corresponds 
precisely to the gauge freedom of the graviton, identifying the theory with perturbative 
general relativity \cite{aa}, \cite{bb}. The concept of parallel transport leads to string-like 
constructions.    

In the following section we consider how gravitation at low energy may be represented 
covariantly by a field on a flat Lorentzian spacetime. This is accomplished by a 
three-index tensor. Then in Section 3 it is shown that the gravitational field is 
determined by the potential by means of Lorentz covariance and symmetry
requirements. The gravitational potential $A_{\mu \nu}$, is shown to be a symmetric, 
two-index tensor field. The next section justifies the hypothesized existence of the
gravitational potential. The wave equation is applied to
the potential of a single point mass, analogously to the Li\'{e}nard-Wiechert 
potential in classical electrodynamics. The solution is used to determine the 
gravitational field tensor and in the static, low-energy limit one recovers 
the Newtonian theory. The final section explores the geometry associated to second-order
covariant differentiation and its connections to string theory.

The gravitational field tensor possesses gauge invariance given 
by a second-order derivative on a line bundle. The concept of second-order covariant 
differentiation is defined in analogy to the first-order case appearing in Yang-Mills theory
and requires the involvement of an auxillary first-order covariant derivative defined 
by a vector field $A_{\mu}$. 
The related geometry lives on the manifold of strings embedded into the spacetime
${\cal M}$, which describes the underlying space for string second quantization \cite{ll}.
It is shown that ${\cal M}$ must be flat and the second-order covariant 
derivative defines a fluctuation from this flat background. Thus the theory restricts
spacetime manifolds to those that admit a flat Lorentian metric. 
Therefore the notion of a flat background from which gravitation propagates is not 
simply a mathematical convenience but a necessary mathematical consequence of the 
hypotheses.

The second-order covariant derivative  reveals the 
composite structure of the graviton in terms of the gravitational potential $A_{\mu \nu}$
and the vector field $A_{\mu}$. The vector field defines the gauge transformations
of the graviton and consequently the hypothesized gravitational potential is seen to
represent the graviton in an arbirary choice of gauge. 
The related geometry is based upon a worldsheet in 
spacetime rather than a worldline, hence its association with string-like objects. 
The symmetric tensor part of the second-order covariant derivative
replicates the bosonic NS-NS fields and couples to the string in a similar manner. 
Parallel translation is carried out over a surface rather than along a line and operates 
on the direct sum of two copies of the string Hilbert space of states.

\section{The Gravitational Field Tensor}

On a flat Lorentzian manifold, a covariant tensor representing
the gravitational field is constructed. This tensor field defines an 
affine connection whose 
geodesics determine the path followed by a particle.  

Let $Q$ be a particle with rest mass $M_{0}$ and 4-position $q(t)=(t,{\bf q}(t))$
on a flat Lorentzian manifold ${\cal M}$ with coordinates $x=(t,{\bf x})$. 
The metric is given by $\eta_{\mu \nu} = diag(+1,-1,-1,-1)$ and $c=1$ in what follows.
Retarded time $t_{0}=t_{0}(x)$ is defined by $t-t_{0}=|{\bf x}-{\bf q}(t_{0})|$; 
gravity is taken to propagate at the speed of light.
The gravitational field of $Q$ will be constructed from the following covariant 
expressions depending upon at most second-order derivatives of $q^{\mu}$: \\
$C^{\mu}(x)$, the lightlike vector field defined by 
$C^{\mu}(x)=x^{\mu}-q^{\mu}(t_{0})$, \\
$U^{\mu}(x) = \frac{dq^{\mu}}{d\tau}(t_{0})$, where $\tau$ is proper time as
measured by $Q$, \\
$\dot{U}^{\mu}(x)=\frac{d^{2}q^{\mu}}{d \tau^{2}} (t_{0})$,\\
$X=C^{\mu} U_{\mu}$, \\
$Y=C^{\mu} \dot{U}_{\mu}$, and \\
$Z=\dot{U}^{\mu}\dot{U}_{\mu}$.

Define the three-index Lorentz-covariant tensor field 
\begin{equation}
H=-GM_{0}X^{-3}C \wedge U\otimes U,   
\label{eq:gravity}
\end{equation}
where the wedge product is given by $C\wedge U=C\otimes U- U\otimes C$ and
$G$ is the gravitational constant.
It is convenient to consider a symmetrized form of $H$:
\begin{equation}
F^{\mu}_{\hspace{0.1in} \nu \lambda} = \frac{1}{2}(H^{\mu}_{\hspace{0.1in} \nu \lambda}
 + H^{\mu}_{\hspace{0.1in} \lambda \nu}).
\label{eq:gravitysym}
\end{equation}  
An affine connection $\Gamma=\Gamma^{\mu}_{\nu \lambda}$
 on the manifold ${\cal M}$ may be defined by 
\begin{equation}
\Gamma=\nabla_{0}-F,
\label{eq:connection}
\end{equation}
where $\nabla_{0}$ denotes the flat Lorentzian connection.

Consider the geodesic equations 
\begin{equation}
\frac{d^{2}x^{\mu}}{ds^{2}} + 
\Gamma^{\mu}_{\nu \lambda}\frac{dx^{\nu}}{ds}\frac{dx^{\lambda}}{ds}=0.
\label{eq:motion}
\end{equation}
Since $F_{(\mu \nu \lambda)}=0$, 
every timelike solution of (\ref{eq:motion}) can be parametrized 
so that $s$ denotes proper time for $x^{\mu}$ with respect to the Lorentzian metric. 
If $Q$ has uniform velocity then in a Lorentz frame where it is at rest at the origin,
the geodesic equations (\ref{eq:motion}) are equivalent to
the Newtonian force law describing the motion $x^{\mu} = x^{\mu}(t)$
of a ponderomotive particle under the influence of the gravitational field of $Q$:
\begin{equation}
{\bf F}=-GM_{0}m{\bf x}/r^{3}, 
\end{equation}
where $r=\sqrt{{\bf x}\cdot {\bf x}}$ and $m$ is relativistic mass.
Therefore, at low energy $F_{\mu \nu \lambda}$
defines the gravitational field of a nonaccelerating point mass in covariant form.  

\section{The Gravitational Potential}

The electrodynamic field tensor is determined locally by a 1-form potential. Analogously,
we hypothesize that to the gravitational field tensor $F$ is associated a  
potential $A$. The latter must be a two-index tensor field and the relationship given by the 
ansatz
\begin{equation}
F_{\mu \nu \lambda} = 
a\partial_{\mu}A_{\nu \lambda} +
b\partial_{\mu}A_{\lambda \nu} +
c\partial_{\nu}A_{\lambda \mu} +
d\partial_{\nu}A_{\mu \lambda} +
e\partial_{\lambda}A_{\mu \nu} +
f\partial_{\lambda}A_{\nu \mu}, 
\label{eq:ansatz}
\end{equation}
for constants $a,b,c,d,e$ and $f$. Symmetry considerations will 
determine all constants appearing in (\ref{eq:ansatz}) up to an overall 
constant multiple which may be absorbed into $A$. First we show that $A$ must be a 
symmetric tensor field. 

In order to ensure that the equations of motion (\ref{eq:motion}) do not involve derivatives 
of $q^{\mu}$ higher than second order, $A_{\mu \nu}$ 
can depend upon at most first-order derivatives.
That is, $A_{\mu \nu}$ must be expressible in terms of $X$, $U^{\mu}$ and $C^{\mu}$ only:
\begin{equation}
A_{\mu \nu} = a(X)U_{\mu}U_{\nu}+b(X)U_{\mu}C_{\nu} + c(X)C_{\mu}U_{\nu} +d(X)C_{\mu}C_{\nu},
\label{eq:unsymmetricansatz}
\end{equation} 
where $a,b,c$ and $d$ are smooth functions.
In the static case where $Q$ resides at the origin, rotational symmetry requires that
$b=c=d=0$. It follows that $A_{\mu \nu}$  has the form
\begin{equation}
A_{\mu \nu}=a(X)U_{\mu}U_{\nu}.
\label{eq:formofA}
\end{equation}
In particular, $A_{\mu \nu}$ is symmetric.

The symmetry relations 
\begin{eqnarray}
 && F_{\mu \nu \lambda} = F_{\mu \lambda \nu}, \\ 
 && F_{\mu \nu \lambda} +F_{\nu \lambda \mu} +F_{\lambda \mu \nu} =0,  
			\hspace{0.1in} and \\
 && A_{\mu \nu} = A_{\nu \mu} 
\end{eqnarray}
yield
\begin{equation}
F_{\mu \nu \lambda} = \partial_{\mu}A_{\nu \lambda} -
\frac{1}{2}(\partial_{\nu}A_{\mu \lambda} + \partial_{\lambda}A_{\mu \nu} )
\label{eq:Ffield}
\end{equation}
and
\begin{equation}
H_{\mu \nu \lambda} = \partial_{\mu}A_{\nu \lambda} -\partial_{\nu}A_{\mu \lambda}. 
\label{eq:Hfield}
\end{equation}

\section{Consistency}

The theory is now defined by a symmetric potential $A_{\mu \nu}$ which determines the 
gravitational field by means of Equation (\ref{eq:Ffield}). In electrodynamics,
the Li\'{e}nard-Wiechert potential describing the field of a single point charge
solves the wave equation in empty space. Analogously, we shall solve the wave equation for the
potential $A_{\mu \nu}$ of the particle $Q$ given by the form (\ref{eq:formofA}) and 
determine the resultant gravitational field tensor. \\
{\bf Remark:} The results that follow actually hold for a much more general ansatz: 
\begin{eqnarray}
A_{\mu \nu} & = & aC_{\mu}C_{\nu}+bC_{\mu}U_{\nu}+cC_{\mu}\dot{U}_{\nu} \nonumber \\
            & + & dU_{\mu}C_{\nu}+eU_{\mu}U_{\nu}+fU_{\mu}\dot{U}_{\nu} \\
		& + & g\dot{U}_{\mu}C_{\nu}+ h\dot{U}_{\mu}U_{\nu}+
			k\dot{U}_{\mu}\dot{U}_{\nu}, \nonumber 
\end{eqnarray}
for arbitrary smooth functions $a,b,c,d,e,f,g,h$ and $k$ of the scalars $X,Y$ and $Z$, 
with the requirement that the partials $\partial_{\mu}A_{\nu \lambda}$ contain no third-order 
derivatives of $q^{\mu}$.

The equation $\Box A_{\mu \nu} = 0$ in empty space has the solution 
\begin{equation}
A_{\mu \nu} = aX^{-1}U_{\mu}U_{\nu},
\label{eq:solution}
\end{equation}
where $a$ is a constant. 
This may be found with the aid of the following derivatives:
\begin{eqnarray}
t_{0},_{\mu} &=& \gamma X^{-1}C_{\mu}, \hspace{0.1in} where \hspace{0.1in} 
\gamma = [1-(\frac{d{\bf q}}{dt}(t_{0}))^{2}]^{-1/2},  \nonumber \\
C_{\mu},_{\nu}&=&\eta_{\mu \nu} - X^{-1}U_{\mu}C_{\nu}, \nonumber \\
U_{\mu},_{\nu}&=&X^{-1}\dot{U}_{\mu}C_{\nu}, \nonumber \\
\dot{U}_{\mu},_{\nu}&=&X^{-1}\ddot{U}_{\mu}C_{\nu}, \hspace{0.1in} where \hspace{0.1in} 
\ddot{U}_{\mu}(x)= \frac{d^{3}q_{\mu}}{d\tau^{3}}(t_{0}), \nonumber  \\
X,_{\nu}&=&U_{\nu}+X^{-1}(Y-1)C_{\nu}, \nonumber \\
Y,_{\nu}&=& \dot{U}_{\nu}+X^{-1}C^{\mu}\ddot{U}_{\mu}C_{\nu}, \hspace{0.1in} and \nonumber \\
Z,_{\nu}&=&2X^{-1}\dot{U}^{\mu}\ddot{U}_{\mu}C_{\nu}. 
\end{eqnarray}

There are two observations to be made.\\
(i) \hspace{0.05in} In the case where $Q$ is at rest at the origin,
$A_{\mu \nu}=\frac{a}{r}\delta_{o \mu}\delta_{0 \nu}$. 
Therefore the $A_{00}$ component corresponds to the Newtonian potential when $a$ is defined to
be $a=-GM_{0}$. \\
(ii) The gravitational field $H$ determined from (\ref{eq:solution}) is
\begin{equation}
H=-GM_{0}X^{-3}[C\wedge (U+X \dot{U})\otimes U + C \wedge U \otimes (-YU+X \dot{U})].
\end{equation}
Restricting to the case where $Q$ has uniform motion gives 
\begin{equation}
 H = -GM_{0}X^{-3}C\wedge U\otimes U. 
\end{equation}
Remarkably, this is the covariant gravitational field
tensor given in (\ref{eq:gravity}); this indicates there is something correct about 
the approach here taken. 
Hence the wave equation applied to the potential in empty space 
leads to Newtonian gravitation in the static limit. This provides justification for the 
hypothesized existence of the gravitational potential. In fact, we shall see in Section 5 
that the potential represents the graviton.

\section{Gauge Invariance and Geometry}

The gravitational field tensor $F$ does not uniquely determine the potential $A$. Rather,
$A$ has the gauge freedom
\begin{equation}
A_{\mu \nu} \rightarrow A_{\mu \nu} + \partial_{\mu} \partial_{\nu}\Lambda,
\label{eq:gauge}
\end{equation}
where $\Lambda$ is an arbitrary smooth function. This differs from gauge theories based upon 
differential forms in that it involves a second-order derivative. 
As gauge theory is intimately tied to differential geometry the associated 
geometric constructions will be considered next. This will lead to the identification of
the gravitational potential $A_{\mu \nu}$ with the graviton and the interpretation of
the gravitational field theory described in Sections 2-4 as linearized general relativity.
Furthermore, string-like entities will appear as the natural generalization, in the second-order
formalism, of geometric objects in conventional (first-order) differential geometry. The 
approach we will follow shall be to define concepts in analogy to the first-order case. 
We begin with the notion of second-order covariant differentiation.  

Let ${\cal M}$ denote a Lorentzian manifold.
A second-order covariant derivative on a vector bundle $E\rightarrow {\cal M}$ is a linear map 
\begin{equation}
 D^{(2)}: \Omega^{0}(E) \rightarrow 
{\cal T}^{2}({\cal M}) \otimes_{ \Omega^{0}({\cal M})} \Omega^{0} (E),
\label{eq:covder}
\end{equation}
which satisfies a Leibniz rule involving an auxillary first-order covariant
derivative $D^{(1)}$ on $E\oplus T{\cal M}$:
\begin{equation}
D^{(2)}(f\xi)=D^{(1)}(df) \otimes \xi
 + S(df \otimes D^{(1)}\xi ) 
 + fD^{(2)}\xi.
\label{eq:Leibniz}
\end{equation}
${\cal T}^{k}({\cal M})= \Gamma^{\infty}(T^{0}_{k}({\cal M}))$ and
$S:  {\cal T}^{2}({\cal M}) \otimes_{ \Omega^{0}({\cal M})} \Omega^{0} (E) \rightarrow 
{\cal T}^{2}({\cal M}) \otimes_{ \Omega^{0}({\cal M})} \Omega^{0} (E) $ 
is the linear map defined by 
\begin{equation}
S( \theta ^{1}\otimes \theta^{2} \otimes Y) = 
    (\theta^{1}\otimes \theta^{2}+\theta^{2}\otimes\theta^{1})\otimes Y,
\end{equation}
for $Y \in \Omega^{0}(E)$ and $\theta^{1},\theta^{2} \in {\cal T}^{1}({\cal M})$.
Equation (\ref{eq:Leibniz}) is chosen so as to mimic the Leibniz rule for 
the square $D^{2}$ of a (first-order) covariant derivative. The auxillary first-order
covariant derivative is a necessary ingredient since second-order differentiation
applied to a product of a function and a vector field will always produce terms
differentiated to first-order only.
$D^{(1)}$ is defined to be the direct sum of two connections 
\begin{equation} D^{(1)} = D^{(1)}_{E}\oplus D^{(1)}_{{\cal M}}, 
\end{equation}
where $D^{(1)}_{E}$ denotes a connection on $E$ and $D^{(1)}_{{\cal M}}$ is the 
Levi-Civita connection on ${\cal M}$. $D^{(1)}_{E}$ is arbitary and shall be
seen to correspond to gauge freedom of the graviton.
Latin indices $i,j,k$ will indicate coordinates with respect to a local frame $\xi_{i}$ for $E$
and Greek indices $\mu, \nu, \lambda$ will represent the spacetime coordinates for ${\cal M}$: 
\begin{eqnarray}
D^{(1)}_{E}(\xi_{j}) &=& (1/G)A^{i}_{\mu j}dx^{\mu} \otimes \xi_{i} \hspace{0.1in} and \\
D^{(1)}_{{\cal M}}(dx^{\mu}) &=& -A^{\mu}_{\nu \lambda}dx^{\nu} \otimes dx^{\lambda}. 
\end{eqnarray}
$D^{(2)}$ determines a tensor field  
$K=K^{i}_{\hspace{0.05in} j \mu \nu}$ by
\begin{equation} D^{(2)}=(D^{(1)})^{2}+(1/G)K. 
\end{equation}

The notion of parallel transport on $E$ for $D^{(2)}$ is sought 
in a manner analogous to that for a 
(first-order) covariant derivative
along a curve $X^{\mu}:[0,1] \rightarrow {\cal M}$:
\begin{equation}
[f^{i},_{\mu}+(1/G)A^{i}_{\mu j} f^{j}]|_{X^{\rho}(t)} \dot{X}^{\mu}=0. 
\label{eq:firstorder}
\end{equation}
To couple the spacetime variables $X^{\mu}$ to the second-order covariant
derivative we separate components from $D^{(2)}$ that are 
tensors with respect to the local spacetime symmetries of ${\cal M}$ and 
with respect to the gauge transformations associated to $E$. Furthermore,
the coupling of a component to $X^{\mu}$ should be a Lie algebra-valued scalar with respect to 
reparameterization. The salient feature as concerns the first-order case is that
(\ref{eq:firstorder}) may be
expressed as a set of ordinary differential equations for $u^{i}(t)=f^{i}(X^{\rho}(t)):$
\begin{equation} \dot{u}^{i} + (1/G)A^{i}_{\mu j}\dot{X}^{\mu} u^{j} =0.
\end{equation}
The requirement of corresponding differential equations in the second-order case will place
restrictions upon the spacetime manifold ${\cal M}$. 

Consider first the $(D^{(1)})^{2}$ part of $D^{(2)}$. $(D^{(1)})^{2}$ decomposes into 
symmetric and anti-symmetric parts:
\begin{equation}(D^{(1)})^{2} = (D^{(1)})^{2}_{sym} + \frac{1}{2}R, 
\end{equation}
where $R=R^{i}_{\hspace{0.05in} j \mu \nu}$ is the curvature for $D^{(1)}_{E}$.
Due to anti-symmetry in the $\mu$ and $\nu$ indices, $R^{i}_{\hspace{0.05in} j \mu \nu}$ 
cannot couple to the vector field $\dot{X}^{\mu}$, defined along a worldline, to produce
a non-zero Lie algebra-valued scalar. 
Therefore the coupling of $X^{\mu}$ to the anti-symmetric part must be of the form 
\begin{equation}
\frac{1}{2\sqrt{|h|}} 
\epsilon^{a b} R^{i}_{\hspace{0.05in} j \mu \nu} X^{\mu},_{a} X^{\nu},_{b},
\end{equation}
where $X^{\mu}: \Sigma \rightarrow {\cal M}$ 
maps a surface $\Sigma$ into ${\cal M}$, $\epsilon^{a b}$ is  the 
anti-symmetric pseudo-tensor
and $h$ denotes the determinant of a metric $h_{a b}$ on $\Sigma$.
Thus the extra index produced by second-order differentiation demands a geometry
based upon a worldsheet in place of a worldline; hence the appearance of strings.

The coupling for the symmetric part of $(D^{(1)})^{2}(\xi)$ is
\begin{equation}
h^{a b}(D^{(1)})^{2}(\xi)_{\mu \nu} X^{\mu},_{a}X^{\nu},_{b}.
\label{eq:symmetricpart}
\end{equation}
In coordinates with $u^{i}(\sigma^{1},\sigma^{2})=f^{i}(X^{\rho}(\sigma^{1},\sigma^{2}))$,
Equation (\ref{eq:symmetricpart}) contains two terms that involve $f^{i}$:
\begin{eqnarray} 
&& f^{i},_{\mu} X^{\mu},_{a b}h^{a b}  \hspace{0.2in} and \\
&& f^{i},_{\mu} A^{\mu}_{\nu \lambda} X^{\nu},_{a} X^{\lambda},_{b}h^{a b}.  
\end{eqnarray}

In order for (\ref{eq:symmetricpart}) to describe partial differential equations 
for $u^{i}$, in suitable coordinates, these two terms must vanish:\\
(i)  \hspace{0.05in}$X^{\mu}$ satisfies the equation $X^{\mu},_{a b}h^{a b} = 0$, and \\
(ii) there exist coordinates on ${\cal M}$ such that $A^{\mu}_{\nu \lambda} =0$. That is,
${\cal M}$ is a flat Lorentzian manifold. This places a topological restriction
upon the possible spacetime manifolds. In particular, the total Chern class of
${\cal M}$ must be zero.

Next, consider the $K$ term. It suffices to restrict the discussion to a line bundle $E$;
henceforth $A^{i}_{\mu j}$ shall be denoted by $A_{\mu}$ and $R^{i}_{\hspace{0.05in}j\mu \nu}$
by $R_{\mu \nu}$.
With respect to the flat background 
$\eta_{\mu \nu}$ on ${\cal M}$, $K$ decomposes into 
three irreducible components:
\begin{equation} 
K_{\mu \nu} = \chi_{\mu \nu} + B_{\mu \nu} + \phi \eta_{\mu \nu},
\label{eq:components}
\end{equation}
where $\chi_{\mu \nu}$
is symmetric and traceless, $B_{\mu \nu}$ is skew-symmetric and $\phi$ is a scalar. 
$\chi_{\mu \nu}$ may be interpreted as the graviton field, describing a perturbation 
$g_{\mu \nu}=\eta_{\mu \nu} + \chi_{\mu \nu}$
from the flat metric $\eta_{\mu \nu}$
and $(1/G)B_{\mu \nu}$ as a type of curvature term. 
Let $R^{(2)}$ be the scalar curvature for $h_{a b}$. 
When the curvatures vanish and $g_{\mu \nu}=\eta_{\mu \nu}$
then parallel transport should reduce to $h^{a b}u,_{a b}=0$,
in suitable coordinates.
This suggests that the scalar $\phi$ 
be multiplied with $R^{(2)}$ in the equation defining parallel transport: 
\begin{equation}
P_{1}+P_{2}+P_{3}+P_{4}=0,
\label{eq:paralleltransport}
\end{equation}
where
\begin{eqnarray}
 P_{1}&=& h^{a b}[u,_{a b} + 	\frac{2}{G}A_{\mu}X^{\mu},_{a}u,_{b}+
	(\frac{1}{G}A_{\mu},_{\nu}+ \frac{1}{G^{2}}A_{\mu}A_{\nu})
        X^{\mu},_{a}X^{\nu},_{b}u], \nonumber \\
 P_{2}&=& \frac{1}{2\sqrt{|h|}} \epsilon^{a b}R_{\mu \nu} (X^{\rho}) 
		X^{\mu},_{a} X^{\nu},_{b}u,  \nonumber \\
 P_{3}&=& \frac{1}{G}[h^{a b}\chi_{\mu \nu}(X^{\rho})+\frac{1}{\sqrt{|h|}}
		\epsilon^{a b} B_{\mu \nu}(X^{\rho}) ]X^{\mu},_{a} X^{\nu},_{b}u,
      \hspace{0.1in} and \nonumber  \\
 P_{4}&=& R^{(2)}\phi(X^{\rho})u. 
\end{eqnarray}
The $K$ field replicates the bosonic NS-NS fields $\chi_{\mu \nu}, B_{\mu \nu}$ and 
$\phi$, and couples to the string $X^{\mu}$ in a similar manner.

The vector field $A_{\mu}$ transforms as 
$A_{\mu} \rightarrow A_{\mu} + \partial_{\mu}\Lambda$ with respect to a gauge 
transformation on the line bundle $E$,
whereas the potential $A_{\mu \nu}$ must transform as
$A_{\mu \nu} \rightarrow A_{\mu \nu} + \partial_{\mu} \partial_{\nu}\Lambda$.
The appearance of the 
term $\frac{1}{2}(\partial_{\mu}A_{\nu}+\partial_{\nu}A_{\mu})$ within the symmetric
part of $D^{(2)}$ motivates the identification
\begin{equation}
A_{\mu \nu} =\chi_{\mu \nu}
 + \frac{1}{2}(\partial_{\mu}A_{\nu}+\partial_{\nu}A_{\mu}).
\label{eq:graviton}
\end{equation}
Therefore, the gauge transformation on $E$ is given by
\begin{eqnarray}
\label{eq:potentialgauge}
A_{\mu \nu} & \rightarrow & A_{\mu \nu} + \partial_{\mu}\partial_{\nu}\Lambda,  \\
A_{\mu} & \rightarrow & A_{\mu} + \partial_{\mu}\Lambda,  \\
f & \rightarrow & exp(-\Lambda/G)f, \hspace{0.1in} and \\
K_{\mu \nu} & \rightarrow & K_{\mu \nu}. 
\end{eqnarray}
Furthermore, it is seen from 
(\ref{eq:graviton}) that $\frac{1}{2}(\partial_{\mu}A_{\nu}+\partial_{\nu}A_{\mu})$
acts as the gauge transformation for the graviton:
\begin{eqnarray}
\label{eq:gravitongauge}
\chi_{\mu \nu} & \rightarrow & \chi_{\mu \nu} +
\frac{1}{2}(\partial_{\mu}A_{\nu}+\partial_{\nu}A_{\mu}).
\end{eqnarray}
The arbitrary nature of the auxillary first-order covariant derivative in the definition
of second-order covariant differentiation naturally describes the gauge freedom of a graviton.
We thus have two separate groups of gauge transformations.

Since a traceless gauge of type (\ref{eq:potentialgauge}),
$\eta^{\mu \nu}A_{\mu \nu}=0$, may be chosen,
$A_{\mu \nu}$ represents the graviton. Therefore, the preceding developments have led
to linearized general relativity. 

Lastly, consider the complexification of $E$. For $h_{a b}$ with metric normal form
$(+1,-1)$, Equation (\ref{eq:paralleltransport}) is hyperbolic and determined by 
initial conditions
\begin{eqnarray}
u(0,\sigma) && 0\leq \sigma \leq \pi, \hspace{0.1in} and \nonumber \\
\label{eq:first}
u,_{\tau}(0,\sigma)  && 0\leq \sigma \leq \pi. 
\label{eqn:second}
\end{eqnarray}
A continuous linear functional $\Psi=\Psi[f]$ of the space of fields $f$
on $[0,\pi]$ can be represented by a continuous complex function $g$ through the inner product
\begin{equation} 
\Psi[f]=\langle g,f \rangle =\int_{0}^{\pi}g^{*}f.
\end{equation} 
These functionals are identified in the Schr\"{o}dinger representation
with the  Hilbert space ${\cal H}(\tau)$ of string states associated to the string
$X^{\mu}(\tau,\sigma)$ at fixed $\tau$. Conversely, the functions 
$u(\tau,\sigma)$ and $u,_{\tau}(\tau,\sigma)$
each define a state in the Hilbert space. Parallel transport
along a surface therefore maps ${\cal H}(\tau_{1})\oplus {\cal H}(\tau_{1})$ into
${\cal H}(\tau_{2})\oplus {\cal H}(\tau_{2})$.

\end{document}